\documentclass[loatfix]{aastex61}
\pdfoutput=1 
\usepackage{amsmath,amstext}
\usepackage[T1]{fontenc}
\usepackage{apjfonts} 
\usepackage[figure,figure*]{hypcap}
\usepackage{enumitem}
\usepackage{multirow}
\usepackage{epstopdf}
\usepackage{comment}

\shortauthors{D. Tak et al.}
\begin{document}
\title{Closure relations of Gamma Ray Bursts in high energy emission}
\author{Donggeun Tak}
\affiliation{Department of Physics, University of Maryland, College Park, MD20742, USA; \href{mailto:takdg123@umd.edu}{donggeun.tak@gmail.com} }
\affiliation{NASA Goddard Space Flight Center, Greenbelt, MD 20771, USA}
\author{Nicola Omodei}
\affiliation{W. W. Hansen Experimental Physics Laboratory, Kavli Institute for Particle Astrophysics and Cosmology, Department of Physics and SLAC National Accelerator Laboratory, Stanford University, Stanford, CA 94305, USA}
\author{Z. Lucas Uhm}
\affiliation{NASA Goddard Space Flight Center, Greenbelt, MD 20771, USA}
\affiliation{Korea Astronomy and Space Science Institute, Daejeon 34055, Republic of Korea}
\author{Judith Racusin}
\affiliation{NASA Goddard Space Flight Center, Greenbelt, MD 20771, USA}
\author{Katsuaki Asano}
\affiliation{Institute for Cosmic Ray Research, The University of Tokyo,
5-1-5 Kashiwanoha, Kashiwa, Chiba 277-8582, Japan}
\author{Julie McEnery}
\affiliation{NASA Goddard Space Flight Center, Greenbelt, MD 20771, USA}
\affiliation{Department of Physics, University of Maryland, College Park, MD 20742, USA}
\begin{abstract}
The synchrotron external shock model predicts the evolution of the spectral ($\beta$) and temporal ($\alpha$) indices during the gamma-ray burst (GRB) afterglow for different environmental density profiles, electron spectral indices, electron cooling regimes, and regions of the spectrum. We study the relationship between $\alpha$ and $\beta$, the so-called ``closure relations'' with GRBs detected by \textit{Fermi} Large Area Telescope (\textit{Fermi}-LAT) from 2008 August to 2018 August. The spectral and temporal indices for the > 100 MeV emission from the \textit{Fermi}-LAT as determined in the Second Fermi-LAT Gamma-ray Burst Catalog \citep[2FLGC;][]{2FLGC} are used in this work. We select GRBs whose spectral and temporal indices are well constrained (58 long-duration GRBs and 1 short-duration GRBs) and classify each GRB into the best-matched relation. As a result, we found that a number of GRBs require a very small fraction of the total energy density contained in the magnetic field ($\epsilon_{B}$ $\lesssim$ 10$^{-7}$). The estimated mean and standard deviation of electron spectral index $\mathit{p}$ are 2.40 and 0.44, respectively. The GRBs satisfying a closure relation of the slow cooling tend to have a softer $\mathit{p}$ value compared to those of the fast cooling. Moreover, the Kolmogorov--Smirnov test of the two $\mathit{p}$ distributions from the fast and slow coolings rejects a hypothesis that the two distributions are drawn from the single reference distribution with a significance of 3.2 $\sigma$. Lastly, the uniform density medium is preferred over the medium that decreases like the inverse of distance squared for long-duration GRBs.
\end{abstract}
\keywords{gamma-ray bursts: general}
\section{Introduction} \label{sec:intro}
Since the late 1960s, thousands of gamma-ray bursts (GRBs) have been detected, and observational and theoretical understandings have progressed \citep[for observational and theoretical reviews, see ][]{Zhang2016, Kumar2015}. A GRB is classified into one of two categories depending on its duration and hardness\footnote{the ratio of total counts between two energy bands; e.g., 50--100 keV and 100--300 keV} \citep{Kouveliotou1993}: long GRB ($T$\textsubscript{90}\footnote{the time during which the cumulative background-subtracted counts increase from 5\% to 95\%} $\geq$ 2s) or short GRB ($T$\textsubscript{90} < 2s). Regardless of the category, a GRB shows the two distinct phases: prompt emission and afterglow. The prompt emission is a short intense emission lasting from a fraction of a second to minutes, and the afterglow is commonly defined as a long-lived emission in a broad energy range from radio to X-ray. The afterglow of GRBs is attributed to the external forward shock where the relativistic ejecta from some central engine interacts with the external medium \citep{Rees1992, Paczynski1993, Meszaros1993, Katz1994, Waxman1997a, Waxman1997b, Meszaros1997}.

According to the external forward shock model, accelerated electrons ($\gamma_{e}$ $\geq$ $\gamma_{m}$) in the forward shock are distributed as a power law, $\frac{dn_{e}}{d\gamma_{e}} \propto \gamma_{e}\,^{-p}$ where $\gamma_{e}$ is the Lorentz factor of electrons, $\gamma_{m}$ is the minimum Lorentz factor in order to form a power-law distribution, and $\mathit{p}$ is the electron spectral index. A fraction of the total energy of the jet goes to the electron distribution ($\epsilon_{e}$) and a different fraction to the generation of a magnetic field ($\epsilon_{B}$). The relativistic electrons interact with the magnetic field and emit synchrotron radiation. Electrons with Lorentz factor higher than a specific Lorentz factor ($\gamma_{c}$) cool down rapidly via synchrotron radiation. If $\gamma_m$ is higher than $\gamma_c$, all electrons cool down rapidly via synchrotron radiation (fast cooling). Inversely, if $\gamma_m$ is lower than $\gamma_c$, electrons with $\gamma_{e}$ $>$ $\gamma_{c}$ can cool down (slow cooling). The spectrum and corresponding light curve from this process, therefore, depend on the relative position of $\nu_{m}$ and $\nu_{c}$ and are described as a series of broken power laws \citep{Sari1998} or smoothly broken power laws \citep{Granot2002}, where $\nu_m$ and $\nu_c$ are the characteristic synchrotron frequencies for the electrons with Lorentz factors $\gamma_m$ and $\gamma_c$, respectively. The temporal ($\alpha$) and spectral ($\beta$) indices in convention of $F_{\nu}$ $\propto$ t$^{-\alpha} \nu^{-\beta}$ are described as a function of the electron spectral index $\mathit{p}$, so that a correlation between $\beta$ and $\alpha$ exists, the so-called closure relation. \cite{Sari1998} derived a set of relations with a uniform density profile ($n$ = constant), and \cite{Chevalier2000} expanded the set of relations in the context of radiation developed in the wind environment where surrounding material density (n) decays as a function of radius ($r$), $n \propto r^{-2}$. The set of the standard closure relations is derived from the combination of three physical properties: (i) a density profile of surrounding materials, (ii) an electron spectral index $\mathit{p}$, and (iii) the electron cooling regime. 

The closure relations have been used as a method for testing the external forward shock model in the X-ray and optical bands \citep[e.g.,][]{Willingale2007,Racusin2009,Wang2015, Fukushima2017}. \cite{Willingale2007} tested the standard closure relations by comparing them with X-ray afterglows detected by the \textit{Neil Gehrels Swift Observatory }\citep[hereafter \textit{Swift};][]{Gehrels2004}. They found that half of their sample satisfies the standard closure relations, but other half needs an alternative model to explain the data. \cite{Racusin2009} performed a systematic study on \textit{Swift} X-ray GRB afterglows with an extensive set of closure relations, and identified the existence of jet breaks \citep{Rhoads1999}, a sudden steepening in a light curve when the beaming angle exceeds the physical collimation angle, in many GRBs. \cite{Wang2015} tested the effectiveness of the external shock model with a set of closure relations by using a large sample of GRBs with both X-ray and optical afterglow data, and concluded that the external forward shock model can account for at least half of GRB afterglows. Since the standard closure relations are not enough to explain the observed, complicated evolution of afterglows, alternative models and corresponding closure relations have been explored \citep[e.g.,][]{Meszaros1998,Sari1999,Chevalier2000,Dai2001,Zhang2004,Zhang2006}.

The \textit{Fermi Gamma Ray Space Telescope} (hereafter \textit{Fermi}) has unveiled high-energy features of GRBs with two instruments, the Gamma-ray Burst Monitor \citep[\textit{Fermi}-GBM; 8 keV--40 MeV; ][]{GBM} and the Large Area Telescope \citep[\textit{Fermi}-LAT; 100 MeV--300 GeV; ][]{LAT}. The energy spectra in the keV--MeV energy band are commonly described by the Band function \citep{Band1993} or a power law with an exponential cutoff. The high-energy spectra of the \textit{Fermi}-LAT energy band ($>$100 MeV) are sometimes not explained by extrapolation of the keV--MeV band, and an additional spectral component usually shaped by a power law is required to fit the observed \textit{Fermi}-LAT data successfully \citep[e.g.,][]{Abdo2009a, Guiriec2010, Ackermann2010, Ackermann2011, Arimoto2016}. The \textit{Fermi}-LAT GeV emission lasts longer than that of keV--MeV emission detected by the \textit{Fermi}-GBM, the so-called ``GeV extended emission'' \citep[e.g.,][]{Ackermann2014}. Interestingly, the light curve of this GeV extended emission shows a power-law decay\citep{Ghisellini2010, Catalog2013(LAT)}, similar to the canonical X-ray afterglow light curve \citep{Zhang2006,Nousek2006}. The analogy between the GeV extended emission and the canonical X-ray afterglow leads to interpretation of the GeV extended emission as emission from the external forward shock \citep[e.g.,][]{Ghirlanda2010, DePasquale2010, Nava2014}. Although several authors show concerns for this approach \citep[e.g.,][]{Maxham2011,ZhangBB2011}, in many studies, the temporal and spectral properties of the GeV extended emission are consistent with the external forward shock model \citep{Kumar2009, Kumar2010, Ghisellini2010, Beniamini2015, Panaitescu2017, Gompertz2018, Tak2019}.

In this paper, we perform a systematic study with high-energy emission of 186 GRBs observed by \textit{Fermi}-LAT from 2008 August to 2018 August \citep{2FLGC}. Out of the 186 GRBs in the catalog, 59 GRBs fulfill our selection criteria and are used. We test the external forward shock model assuming adiabatic hydrodynamic evolution by comparing the properties of the selected GRBs with a set of the standard closure relations. A uniform density profile of the interstellar medium (ISM; $n$ = constant) and a wind profile ($n(r)$ $\propto r^{-2}$) are tested. In addition to the circumburst condition, three cooling regimes are considered; $\nu$ > $\nu_{m}$ and $\nu_{c}$, $\nu_{m}$ < $\nu$ < $\nu_{c}$, and $\nu_{c}$ < $\nu$ < $\nu_{m}$. Since the synchrotron self-absorption process is relevant to emission well below infrared \citep[e.g.,][]{Katz1994}, we neglect the closure relations related to the self-absorption regime. We also assume that the temporal indices follow evolution before the jet break. We find the best-matched closure relation for each GRB, and discuss implications of the physical conditions of GRBs deduced from the best-matched closure relation. Data preparation is presented in section~\ref{sec:data}. In section~\ref{sec:method}, we describe a set of closure relations and a method for classifying GRBs into the most probable closure relation. In section~\ref{sec:result} and section~\ref{sec:discuss}, we describe the classification result and explore the implication of the results, respectively. We summarize and reach a conclusion in section~\ref{sec:conclusion}.

\section{Data preparation} \label{sec:data}
We define the \textit{Fermi}-LAT extended emission as emission after \textit{Fermi}-GBM T$_{90}$ measured in the energy range from 50 to 300 keV. The light curve and energy spectrum of the \textit{Fermi}-LAT extended emission are analyzed in the Second Fermi-LAT Gamma-Ray Burst Catalog (2FLGC), and we adopt the values of temporal and spectral indices from the catalog.

In the catalog, a light curve of a GRB extended emission consisting of at least three flux points is fitted with a simple power law and a broken power law. Since a broken power law is never significantly better than a simple power law, we adopt a temporal index from the simple power law. We impose a criterion on the error size of a GRB temporal index to remove a GRB with a large error, resulting in a meaningless classification. The error size of the temporal index is required to be less than 1/2 where two typical closure relations, $\alpha$ = (3$\beta$ - 1)/2 and $\alpha$ = 3$\beta$/2, are differentiated by the value of 1/2 when its spectral index is the same.

The energy spectrum of the GRB extended emission is fitted with a simple power law in the energy range from 100 MeV to 100 GeV. Similarly to the cut on the error of the temporal index, we also require the error on the spectral index to be less than 1/3. The difference in the error-size criteria results from the consideration of the maximum slope of a closure relation ($\alpha$ $\sim$ 3$\beta$/2). After these procedures, 59 out of 186 GRBs are selected (58 long and 1 short GRBs).
\begin{table*}[t]
	\centering 
	\caption{The set of closure relations fit in this study.}
	\begin{tabular}{c c c c c c c c c}
    \hline\hline
 & \multicolumn{3}{c}{Condition} & &\multicolumn{3}{c}{Closure Relation  }\\\cline{2-4}\cline{6-8}
Class&$\mathit{p}$&Cooling regime &Environment& & $\beta$($\mathit{p}$) & $\alpha$($\mathit{p}$) & $\alpha$($\beta$) \\\hline
CR1&-&$\nu\textsubscript{c}$ < $\nu$ < $\nu\textsubscript{m}$&ISM/wind& &1/2&1/4&-\\
CR2&$\mathit{p}$ > 2&$\nu$ > $\nu\textsubscript{m}$, $\nu\textsubscript{c}$&ISM/wind& &$\mathit{p}$/2&(3$\mathit{p}$-2)/4& (3$\beta$-1)/2\\
CR3&$\mathit{p}$ > 2&$\nu\textsubscript{m}$ < $\nu$ < $\nu\textsubscript{c}$&ISM& &($\mathit{p}$-1)/2&3($\mathit{p}$-1)/4& 3$\beta$/2\\
CR4&$\mathit{p}$ > 2&$\nu\textsubscript{m}$ < $\nu$ < $\nu\textsubscript{c}$&wind& &($\mathit{p}$-1)/2&(3$\mathit{p}$-1)/4& (3$\beta$+1)/2\\
CR5&1 < $\mathit{p}$ < 2&$\nu$ > $\nu\textsubscript{m}$, $\nu\textsubscript{c}$&ISM& &$\mathit{p}$/2&(3$\mathit{p}$+10)/16& (3$\beta$+5)/8\\
CR6&1 < $\mathit{p}$ < 2&$\nu\textsubscript{m}$ < $\nu$ < $\nu\textsubscript{c}$&ISM& &($\mathit{p}$-1)/2&3($\mathit{p}$+2)/16& (6$\beta$+9)/16\\
CR7&1 < $\mathit{p}$ < 2&$\nu$ > $\nu\textsubscript{m}$, $\nu\textsubscript{c}$& wind& &$\mathit{p}$/2&($\mathit{p}$+6)/8& ($\beta$+3)/4\\
CR8&1 < $\mathit{p}$ < 2&$\nu\textsubscript{m}$ < $\nu$ < $\nu\textsubscript{c}$&wind& &($\mathit{p}$-1)/2&($\mathit{p}$+8)/8& (2$\beta$+9)/8\\
	\hline\hline
		\end{tabular}
        
\label{tab:cr}
\end{table*}

\section{Classification Method} \label{sec:method}
In Table~\ref{tab:cr}, we summarize the standard closure relations \citep{Gao2013}, eight in total. For the set of spectral ($\beta$) and temporal ($\alpha$) index pairs satisfying our selection criteria, we perform a statistical analysis to find which closure relation best fits each pair.

According to Bayes' theorem, the probability of a closure relation given observed $\beta$ and $\alpha$ is
\begin{equation}
\textit{P}(CR \mid \alpha, \beta) = \frac{\textit{P}(\alpha, \beta \mid CR)\,\textit{P}(CR)}{\textit{P}(\alpha, \beta)}.
\end{equation}
For any choice of two closure relations, the Bayes factor $K$ can be computed for comparing which model better describes the observed data
\begin{equation}
K_{ij} = \frac{\textit{P}(\alpha, \beta \mid CR_{i})}{\textit{P}(\alpha, \beta \mid CR_{j})} = \frac{\textit{P}(CR_{i} \mid \alpha, \beta)\,\textit{P}(CR_{j})}{\textit{P}(CR_{j} \mid \alpha, \beta)\,\textit{P}(CR_{i})}.
\end{equation}
The prior probability, $\textit{P}(CR_{j})$, of each closure relation can be estimated from other information such as observational results at other wavelengths or physical intuition. For example, each closure relation is derived from different ranges of the electron spectral index (either $\mathit{p}$ > 2 or 1 < $\mathit{p}$ < 2). Since an estimated value of the electron spectral index $\mathit{p}$ from other observational results is generally larger than 2, we can impose a prior probability that the closure relations derived from $\mathit{p}$ > 2 are more probable than the others. However, in this work, we made the decision to have the prior probability of each closure relation equal, \textit{P}(CR$_{i}$) = \textit{P}(CR$_{j}$); i.e., all closure relations are equally probable.

According to the external forward shock model, the spectral and temporal indices are a function of the electron spectral index $\mathit{p}$, except for CR1 (see Table~\ref{tab:cr}). Because both $\beta$ and $\alpha$ are a function of $\mathit{p}$, we can convert $\beta$ and $\alpha$ to $\mathit{p}_{\beta}$ and $\mathit{p}_{\alpha}$, respectively. The likelihood \textit{P}($\alpha$, $\beta$ $\mid$ CR$_{i}$) can therefore be estimated by requiring that both $\mathit{p}_{\beta}$ and $\mathit{p}_{\alpha}$ satisfy the constraint of the closure relation and that $\mathit{p}_{\beta}$ and $\mathit{p}_{\alpha}$ are identical. Under the assumption that the uncertainties on $\mathit{p}_{\beta}$ and $\mathit{p}_{\alpha}$ are normally distributed and considering these two requirements, the quantity \textit{P}($\alpha$, $\beta$ $\mid$ CR$_{i}$) can be described as
\begin{equation}
\begin{aligned}
\textit{P}(\alpha, \beta, \mid CR_{i})\, =\, &\textit{P}(\mathit{p}_{\alpha,\,i} = \mathit{p}_{\beta,\,i} \mid CR_{i})\prod_{j\,=\,\alpha,\,\beta}\textit{P}(\mathit{p}_{j,\,i} \subset \mathit{p}_{c} \mid CR_{i}),
\end{aligned}
\end{equation}
where $\mathit{p}_{c}$ refers to the $\mathit{p}$ constraint of the corresponding closure relation. The likelihood \textit{Pr} ($\mathit{p}_{\alpha}$ = $\mathit{p}_{\beta}$ $\mid$ CR$_{i}$) is
\begin{equation}
\textit{P}\,(\mathit{p}_{\alpha,\,i} = \mathit{p}_{\beta,\,i} \,\mid\,CR_{i}) = \frac{1}{\sqrt{2\pi(\sigma_{\mathit{p}_{\alpha,\,i}}^{2}+\sigma_{\mathit{p}_{\beta,\,i}}^{2})}}exp\left[-\frac{(\mathit{p}_{\beta,\,i}\,-\,\mathit{p}_{\alpha,\,i})^2}{2(\sigma_{\mathit{p}_{\alpha,\,i}}^{2}+\sigma_{\mathit{p}_{\beta,\,i}}^{2})}\right],
\end{equation}
where $\sigma_{\mathit{p}_{\alpha,\,i}}$ and $\sigma_{\mathit{p}_{\beta,\,i}}$ are the 1$\sigma$ error of $\mathit{p}_{\alpha,\,i}$ and $\mathit{p}_{\beta,\,i}$, respectively. The probability P($\mathit{p}_{j} \subset \mathit{p}_{c}$ $\mid$ CR$_{i}$) can be calculated by  
\begin{equation}
P(\mathit{p}_{j} \subset \mathit{p}_{c} \mid CR_{i}) \\
=
\begin{cases}
1 & \mbox{if } \mathit{p}_{j} \subset \mathit{p}_{c},\\2 \times
\int\limits_{\text{$\mathit{p}$ in $\mathit{p}_{c}$}}^{} \frac{1}{\sigma_{\mathit{p}_j}\sqrt{2\pi}}exp\left[-\frac{(\mathit{p}\,-\,\mathit{p}_{j})^2}{2\sigma_{\mathit{p}_j}^2}\right] d\mathit{p} &\mbox{otherwise},
\end{cases}
\end{equation}
which is a two-sided probability of the region satisfying the $\mathit{p}$ constraint. 

In the case of CR1, $\beta$ and $\alpha$ are not functions of $\mathit{p}$, and they are independent. Instead, we calculate the likelihood for the data with 
\begin{equation}
\textit{P}(\alpha, \beta \mid CR_{1})=\frac{1}{2\pi\sigma_{\alpha}\sigma_{\beta}}exp\left[{-\frac{(\alpha\,-\,1/4)^2}{2\sigma_{\alpha}^2}-\frac{(\beta\,-\,1/2)^2}{2\sigma_{\beta}^2}}\right]. 
\end{equation}

The best-matched closure relation is determined by comparing the quantity of \textit{P}(CR$_{i}$$\mid$$\alpha$, $\beta$). A model with the highest \textit{P}(CR$_{i}$$\mid$$\alpha$, $\beta$) is a possible best-matched model. However, if the possible best-matched model gives $(\mathit{p}_{\beta,\,\rm best}\,-\,\mathit{p}_{\alpha,\,\rm best})^2/(\sigma_{\mathit{p}_{\alpha,\,\rm best}}^{2}+\sigma_{\mathit{p}_{\beta,\,\rm best}}^{2})$ greater than 1.65, the best-matched closure relation is not assigned; i.e., such GRBs are called "unclassified (UC)".  When the best-matched closure relation is assigned, the Bayes factor against the other closure relations is computed. The smallest Bayes factor, which is given by the best-matched closure relation against the most competitive alternative, can be used for estimating the strength of how much the best-matched closure relation is supported by the data against the others. A Bayes factor smaller than 3 means that the best-matched closure relation is not significantly better than the other closure relations \citep{Robert1995}.

\section{Closure Relation Classification Results} \label{sec:result}
\begin{figure*}[t]
  \centering
  \includegraphics[scale=0.8]{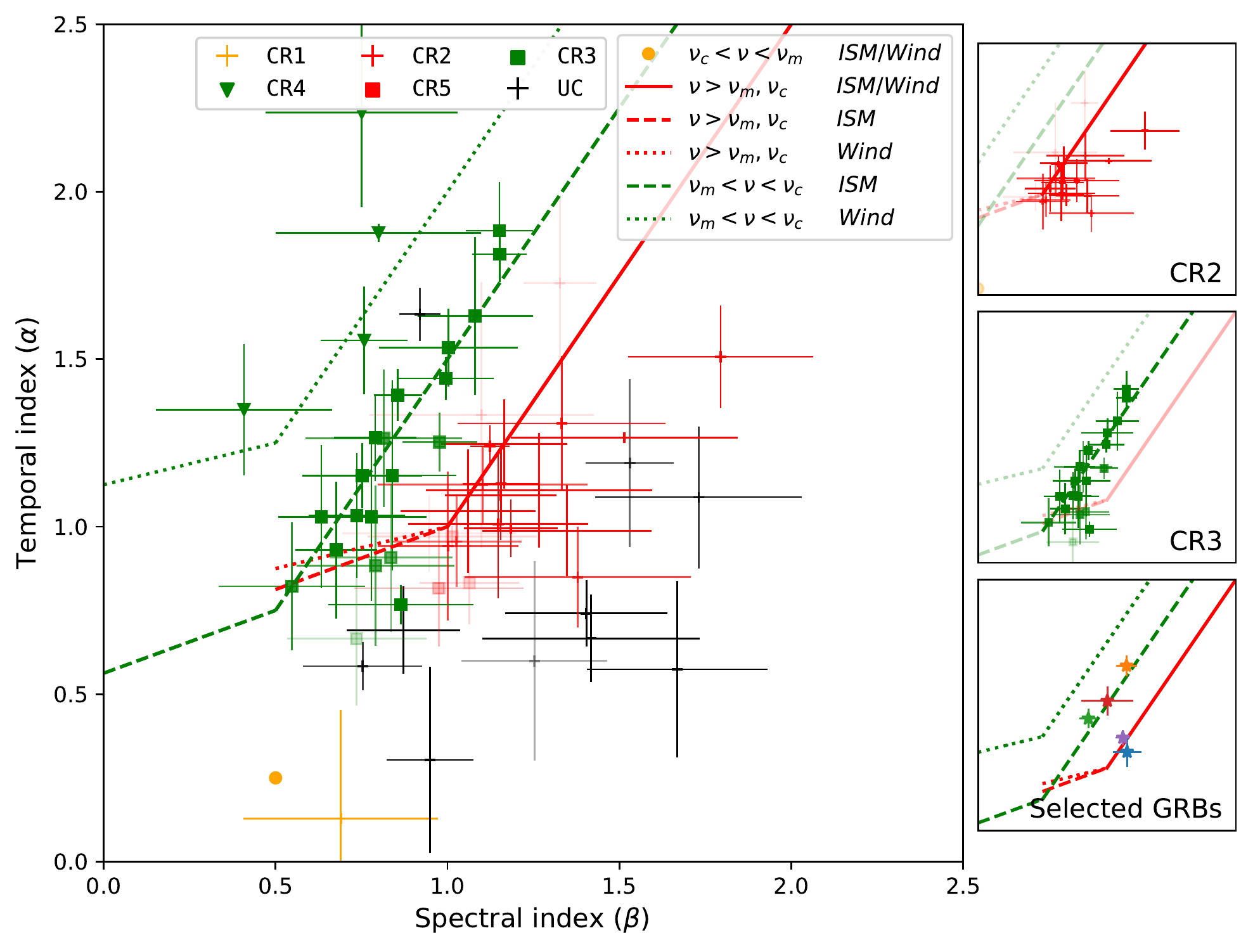}
  \caption{Classification result. The color of data points and lines differentiates a cooling regime; red is used for the fast cooling ($\nu$ > $\nu\textsubscript{m}$ and $\nu\textsubscript{c}$), green for the slow cooling ($\nu\textsubscript{m}$ < $\nu$ < $\nu\textsubscript{c}$), orange for $\nu\textsubscript{c}$ < $\nu$ < $\nu\textsubscript{m}$. Note that for $\nu\textsubscript{c}$ < $\nu$ < $\nu\textsubscript{m}$, the spectral and temporal indices are specifically determined: $\beta$ = 0.5 and $\alpha$ = 0.25. Each set of $\alpha$ and $\beta$ is depicted by color corresponding to the best-matched closure relation, but an unclassified GRB is displayed in black. The closure relation line style is related to the surrounding environment property; a solid line is used for an undetermined environment, a dashed line for ISM, and a dotted line for wind. Two subpanels (top and middle) show classified events for the two most frequently used closure relations, separately. Also, a bottom subpanel shows some selected GRBs, which are broadly studied with multiwavelength observations: GRB 080916C (blue), GRB 090510A (orange), GRB 090926A (green), GRB 110731A (red), and GRB 130427A (purple) (see section~\ref{sec:multi})}.
\label{fig:result}
\end{figure*}

\begin{table*}[t]
	\centering 
	\caption{The number of classified GRBs for each closure relation.}
	\begin{tabular}{*{13}{c}}
    \hline\hline
\multirow{3}{*}{Class}&&$\nu_{c}$ < $\nu$ < $\nu_{m}$&&\multicolumn{2}{c}{$\nu$ > $\nu_{m}$, $\nu_{c}$}&&\multicolumn{2}{c}{$\nu_{m}$ < $\nu$ < $\nu_{c}$}&\multirow{3}{*}{UC\footnote{unclassified}}&\multirow{3}{*}{Total}\\\cline{3-3}\cline{5-6}\cline{8-9}
&&I/W\footnote{ISM/wind}&&I/W&ISM\footnote{1 < \textit{p} < 2\label{ft:p}}&&ISM&wind&&\\
&&CR1&&CR2&CR5&&CR3&CR4&&\\\hline
Long&&1&&20&3&&19&4&11&58\\
Short&&-&&-&-&&1&-&-&1\\\hline
Total&&1&&20&3&&20&4&11&59\\
\hline\hline
		\end{tabular}
\label{tab:result}     
\end{table*}
%

\begin{figure}[t]
  \centering
  \includegraphics[scale=0.8]{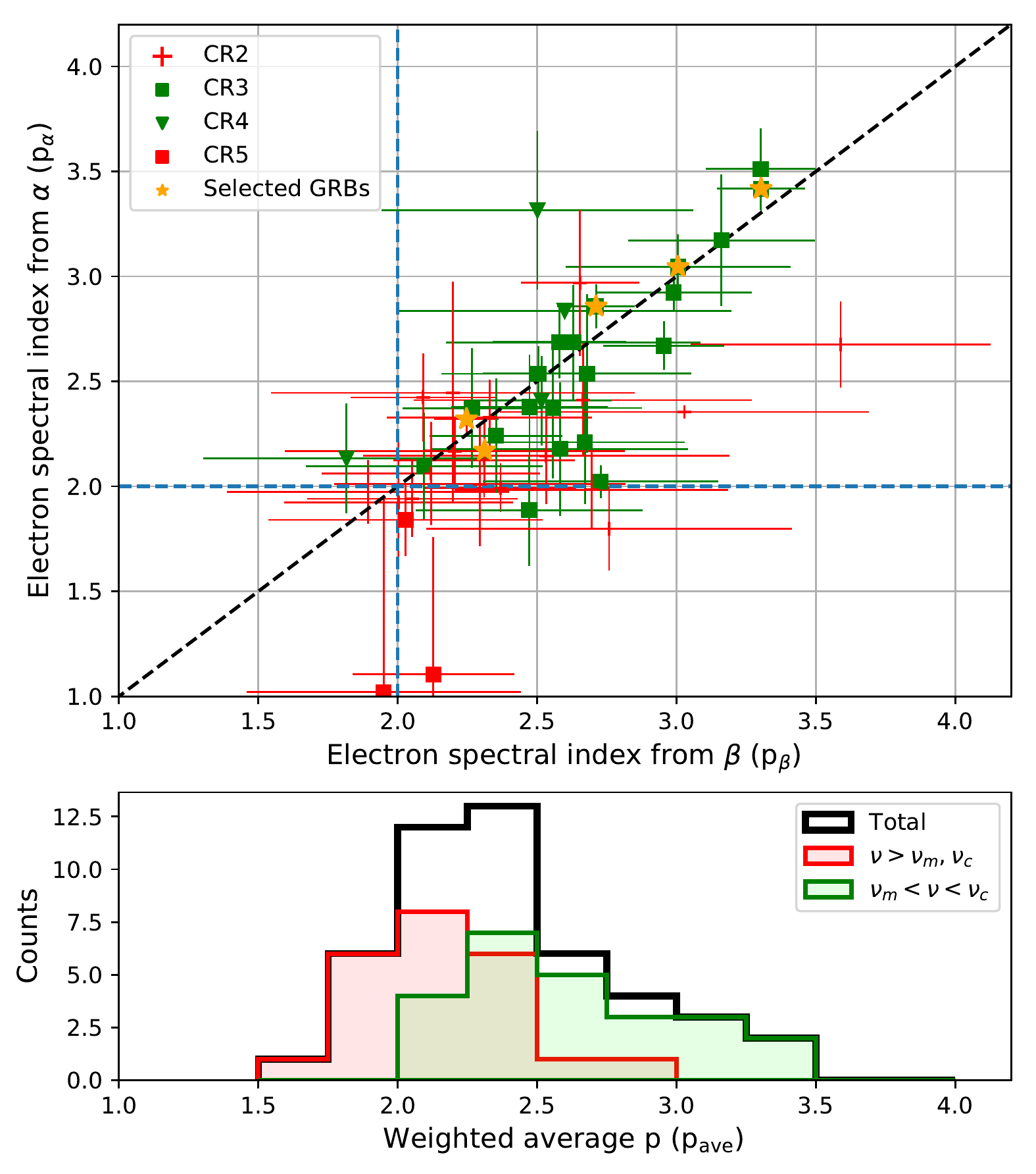}
  \caption{Distribution of the electron spectral index $\mathit{p}$. Upper panel: scatter plot of $\mathit{p}_{\beta}$ versus $\mathit{p}_{\alpha}$ for 48 classified GRBs. Each data point is displayed in red (fast cooling; $\nu$ > $\nu_{m}$, $\nu_{c}$) or green (slow cooling; $\nu_{m}$ < $\nu$ < $\nu_{c}$) depending on the cooling regime of the best-matched closure relation. The Black dashed line indicates the identity line. The selected GRBs, GRB 080916C, GRB 090510A, GRB 090926A, GRB 110731A, and GRB 130427A, are marked with an additional yellow star point. Bottom panel: the distribution of the weighted average of $\mathit{p}_{\beta}$ and $\mathit{p}_{\alpha}$. }
\label{fig:pdist} 
\end{figure}

The classification results are summarized in Figure~\ref{fig:result} and Table~\ref{tab:result}, details for all GRBs are given in the Appendix (Table~\ref{tab:tot}). In total, 48 out of 59 GRBs are classified into one of the standard closure relations, and the others are unclassified. Only 5 of the 8 closure relations, CR1--CR5, are preferred for the classified GRBs. In Figure~\ref{fig:result}, the set of closure relations, and 59 GRBs are plotted. Depending on the value of the Bayes factor compared to the most competing model, the degree of transparency is determined; i.e., the less ambiguity of the classification, the higher the visibility. We note that there are some green points on top of the red dotted and dashed lines. This is because for those closure relations (CR5 and CR6), the error size of $\mathit{p}_{\alpha}$ ($\sigma_{\mathit{p}_{\alpha}}$) is four times larger than the one given by other closure relations (CR2, CR3, and CR4), so that the value of \textit{P}($\alpha$, $\beta$ $\mid$ CR$_{i}$) decreases significantly. Most of the classified events have spectral and temporal indices from 0.5 to 1.5, but the unclassified GRBs tend to have either a soft spectrum ($\beta$ $\sim$ 1.5) or a shallow decay ($\alpha$ < 0.5). Table~\ref{tab:result} shows how many GRBs are sorted into each closure relation. The most frequent classified class is CR3 derived from a cooling condition of $\nu_{m}$ < $\nu$ < $\nu_{c}$ with a uniform density profile (ISM). Also, many GRBs are classified into CR2 for which $\nu$ is higher than both $\nu_{m}$ and $\nu_{c}$ with an undetermined circumburst environment.

We compute $\mathit{p}_{\alpha}$ and $\mathit{p}_{\beta}$ from $\alpha$ and $\beta$, respectively, by applying the conversions of the best-matched closure relation. Figure~\ref{fig:pdist} shows a scatter plot of $\mathit{p}_{\beta}$ versus $\mathit{p}_{\alpha}$ and the distribution of the weighted average of $\mathit{p}_{\beta}$ and $\mathit{p}_{\alpha}$. The red (CR 2 and 5) and green points (CR 3 and 4) are related to the different cooling regime, the fast cooling ($\nu$ > $\nu_{c}$ and $\nu_{m}$) and the slow cooling ($\nu_{m}$ < $\nu$ < $\nu_{c}$), respectively. In the upper panel of Figure~\ref{fig:pdist}, the red and green points are clustered in different areas. Such deviation is also evident in the weighted $\mathit{p}$ distribution of the fast cooling and the slow cooling (see the lower panel of Figure~\ref{fig:pdist}. The detailed discussion on the $\mathit{p}$ value in the two different cooling regimes is in section~\ref{sec:pvalue}.

\section{Discussion} \label{sec:discuss}
\subsection{Environment for \textit{Fermi}-LAT extended emission} \label{sec:environment}
A long GRB is attributed to the core-collapse of a massive star \citep[e.g.,][]{Woosley1993, MacFadyen1999, Woosley2006}, and stellar winds from the massive star form the circumburst density profile as a function of radius \citep[e.g.][]{Chevalier1999}. As the radius from the central engine increases, the density profile is no longer purely described by the stellar winds, but is influenced by the interstellar medium. A short GRB originates from the merger of two compact objects, and thus a uniform density profile with a lower density environment is expected \citep[e.g.,][]{Paczynski1986, Paczynski1991, Rosswog1999, GW170817}.

Among 58 long GRBs, we identify the environment of 26 GRBs, and most of them (22 out of 26) are related to the ISM. Other observational studies reach similar conclusions \citep[e.g,][]{Yost2003,Schulze2011}. \cite{Yost2003} found that all of their whole long GRB sample fits best with a constant density medium from 0.2 to 20 cm\textsuperscript{-3}. Also, \cite{Schulze2011} performed a statistical study with \textit{Swift} GRBs and found that the majority of the long GRBs are compatible with a uniform density profile rather than a wind profile. These results are in tension with the expectation that the progenitors of long GRBs are massive stars whose external density decreases with radius in proportional to radius squared. This implies that a wind profile may not be extended at the radii of the forward external shock \citep{Schulze2011}. Another possible explanation for such conflicted results is that our assumption of the wind density profile is too simple; i.e., the wind profile can deviate from the simple description \cite[e.g.,][]{Eldridge2006}.

In our sample, there is one short GRB (GRB 090510), which is related to the ISM consistent with the theoretical expectation.

In this work, we test the simple external forward shock model where the density profile is either a uniform profile (ISM) or a wind profile. However, the density profile parameter $k$, $n(r)$ $\propto$ r$^{-k}$, is not required to be fixed to 0 or 2 \cite[e.g.,][]{Granot2002, Kouveliotou2013}. Finding an appropriate k value for each GRB is beyond the scope of this work.

\subsection{Cooling regime for \textit{Fermi}-LAT extended emission} \label{sec:cooling}
In this work, we find that 24 GRBs are classified into the slow-cooling condition, $\nu_{m}$ < $\nu$ < $\nu_{c}$ (Table ~\ref{tab:result}). This result implies that for those GRBs $\nu_{c}$ is required to be higher than the \textit{Fermi}-LAT extended emission energy band ($\nu_{\rm LAT}$) in order to satisfy this cooling condition. Since the spectral and temporal indices are computed from the energy band, 100 MeV--10 GeV, $\nu_{c}$ should be at least higher than a frequency equivalent to 100 MeV. 

According to the external forward shock model, the cooling frequency $\nu_{c}$ is described as a function of four parameters: the fraction of the total energy behind the shock in the magnetic field $\epsilon_{B}$, the isotropic energy $E_{\rm iso}$, the circumburst density profile $n$, and the observed time $t_{\rm obs}$. Among the four parameters, the distributions of the isotropic energy and the observed time do not show any dependence on the cooling condition (2FLGC). We take the average value of $E_{\rm iso}$ and $t_{\rm obs}$ from 2FLGC, $E_{\rm iso}$ $\sim$ 10$^{53}$ erg and $t_{\rm obs}$ $\sim$ 10$^{3}$ s. Since most GRBs favor the ISM circumburst condition (section~\ref{sec:environment}), the density profile is assumed to be ISM, $n$ = 1 cm$^{-3}$. Under these reasonable assumptions, $\epsilon_{B}$ has an upper limit \citep{Sari2001},
\begin{equation}
\begin{aligned}
&\epsilon_{B} \lesssim 4.8 \times 10^{-7} \ (1+z)^{-1/3} \ \nu_{c,\ 100\ MeV} {}^{-2/3} \ E_{53} {}^{-1/3} \ n_{\,1} {}^{-2/3} \ t_{3} {}^{-1/3} \ (1+Y)^{-4/3},\\
\end{aligned}
\end{equation}
where $\nu_{c,\ 100\ MeV}$ = $h \nu_{c}$ / 100 MeV, $E_{53}$ = $E_{\rm iso}$ / 10$^{53}$ erg, $n_{\, 1}$ = $n$ / 1 cm$^{-3}$, and $t_{3}$ = $t_{\rm obs}$ / 10$^{3}$ s. Even with $Y \ll 1$, where $Y$ is the parameter of the relative power of the synchrotron self-Compton (SSC) emission compared to the synchrotron emission \citep[e.g.,][]{Kumar2009, Wang2010}, the GRBs classified into the slow-cooling condition require a very small value of $\epsilon_{B}$ ($\lesssim$ 10$^{-7}$). Such a small value of $\epsilon_{B}$, however, does not conflict with other observational studies in the X-ray, optical, and radio bands \citep[e.g.,][]{Santana2014, Kumar2009, Kumar2010, Duran2014, Wang2015, Beniamini2015, Zhang2015}. For example, \cite{Santana2014} performed a systematic study on magnetic fields in GRB afterglows by using mostly \textit{Swift} data. They computed $\epsilon_{B}$ for X-ray and optical afterglows and found that the distributions of $\epsilon_{B}$ for X-ray (optical) have a range of 10\textsuperscript{-8} $\sim$ 10\textsuperscript{-3} (10\textsuperscript{-6} $\sim$ 10\textsuperscript{-3}) and the medians of about few 10\textsuperscript{-5} in both X-ray and optical.  

For GRBs with a very small $\epsilon_{B}$, the contribution from the SSC emission to the GeV emission cannot be ignored, because the relative contribution of the SSC emission depends on the value of $\epsilon_{B}$ ($Y$ $\propto$ $\epsilon_{B}^{-1/2}$, when $Y$ $\gg$ 1) \citep{Panaitescu2000, Sari2001, Nakar2009}. Several studies show that the SSC contribution to the LAT energy band from 100 MeV to a few GeV is small even for Y $\ll$ 1 because of the Klein-Nishina effect  \citep[e.g.,][]{Kumar2009, Wang2010}, but it is possible that the emerging SSC emission can lead spectral and temporal indices deviated from those expected from the synchrotron process. However, we have not found strong evidence of SSC-induced deviation from the standard closure relation at present. To assess the specific contribution of SSC emission, the detailed spectrum and its evolution of the electron distribution should be studied by multiwavelength modeling \citep[e.g.,][]{Beniamini2015}, which is unknown in a systematic way. Instead, the relative change in the spectral and temporal indices due to the contribution of SSC can be explored. \cite{Lemoine2015} studied the SSC spectrum in the very small $\epsilon_{B}$ regime, and calculated the expected evolution of $\beta$ and $\alpha$. With the constant density $n$ = 1 cm$^{-3}$, $\epsilon_{B}$ = 10$^{-5}$, \textit{p} = 2.3, and $t_{\rm obs}$ $\sim$ 10$^{2}$--10$^{3}$, the expected $\beta$ and $\alpha$ in 0.1--10 GeV are $\beta$ $\sim$ 0.75--0.9 and $\alpha$ $\sim$ 1.0, which are different from those without the SSC contribution; $\beta$ = 1.15 and $\alpha$ = 1.225 for the fast cooling, and $\beta$ = 0.65 and $\alpha$ = 0.975 for the slow cooling. This set of $\beta$ and $\alpha$ is located between the slow cooling line (dashed green in Figure~\ref{fig:result}) and fast cooling line (solid red in Figure~\ref{fig:result}). \cite{Fukushima2017} also found that the standard closure relation can deviate due to emerging SSC emission, and such deviation directs to the area between the slow cooling line and the fast cooling line. These studies suggest that for GRBs with a set of $\beta$ and $\alpha$ scattered around such region, the SSC contribution may be important in understanding their evolution of spectrum and light curve, and the standard closure relations may not work properly.

\subsection{Electron spectral index} \label{sec:pvalue}
Both theoretical studies and numerical simulations on an electron spectrum developed in a relativistic shock show that an electron spectral index $\mathit{p}$ has a universal value $\sim$ 2.2 -- 2.4 \citep{Meszaros1998, Kirk2000, Achterberg2001, Spitkovsky2008}. In contrast, the observational studies found that $\mathit{p}$ varies from one GRB to another, and the distribution of $\mathit{p}$ forms a Gaussian function \citep{Shen2006, Starling2008, Curran2009, Curran2010}. In accordance with other observational studies, Figure~\ref{fig:pdist} shows that the weighted averages of $\mathit{p}$ obtained from the classified GRBs are distributed in the range from 1.5 to 3.5, rather than having a universal $\mathit{p}$. To get median and 1 $\sigma$ of the $\mathit{p}$ distribution, we perform a simulation study. For each $\mathit{p}_{\rm ave}$ value, we assume that mean and error of $\mathit{p}_{\rm ave}$ follow a normal distribution, and randomly generate a $\mathit{p}$ value with a probability given by the normal distribution. We gather the simulated $\mathit{p}$ values and make a distribution. The simulated distribution is fitted with a Gaussian function by the maximum likelihood method. We repeat this procedure 10$^{5}$ times. From the 10$^{5}$ set of median and 1 $\sigma$, we estimate median and 1 $\sigma$ of the true $\mathit{p}$ distribution, which are $\mathit{p}$ = 2.40 $\pm$ 0.03 and $\sigma_{\mathit{p}}$ = 0.44 $\pm$ 0.03. This result is consistent with \cite{Curran2010} in which the distribution of $\mathit{p}$ measured from \textit{Swift} GRB X-ray afterglows is well described by a Gaussian function centered at $\mathit{p}$ = 2.36 and having the standard deviation of 0.59.

Next, we test the dependence of $\mathit{p}$ on the electron cooling condition. The $\mathit{p}_{\rm ave}$ distribution of CR2 (fast cooling) and the combined $\mathit{p}_{\rm ave}$ distribution of CR3 and CR4 (slow cooling) are compared. The simulation for these two sets yields that the $\mathit{p}_{\rm ave}$ distribution of the fast cooling has median and 1 $\sigma$ of 2.22 $\pm$ 0.04 and 0.30 $\pm$ 0.04, respectively, and in the case of the slow cooling, median and 1 $\sigma$ are 2.61 $\pm$ 0.04 and 0.43 $^{+0.04}_{-0.03}$, respectively. Since the two distributions are not aligned (see lower panel in Figure~\ref{fig:pdist}), we perform the Kolmogorov--Smirnov (KS) test on them. As a result, the null hypothesis that the two distributions are drawn from the same reference distribution is rejected at 3.2 $\sigma$ (two-sided p-value = 1.2$\times 10^{-3}$). This implies that the slope of the electron spectral distribution is possibly related to the cooling regime.

In addition, we explore the dependence of $\mathit{p}$ on the surrounding environment. We compare CR3 (ISM) and CR4 (wind). In the case of ISM, the median and standard deviation are 2.62 $\pm$ 0.04 and 0.43 $\pm$ 0.04, respectively. For wind, the median and standard deviation are 2.60 $\pm$ 0.11 and 0.40 $\substack{+0.13 \\ -0.11}$, respectively. Since the two results agree within 1 $\sigma$, the dependence of $\mathit{p}$ on surrounding environment is unclear.

\subsection{Comparison with multiwavelength results} \label{sec:multi}
We choose some selected GRBs that show a clear high-energy extended emission observed by \textit{Fermi}-LAT and multi-wavelength afterglow, so that the physical conditions of the external forward shock have been well studied and understood. They are GRB 080916C, GRB 090510A, GRB 090926A, GRB 110731A, and GRB 130427A (see Figure~\ref{fig:result} and Figure~\ref{fig:pdist}). For these GRBs, we compare our results with multiwavelength results to check consistency.

GRB 080916C (blue in the subpanel of Figure~\ref{fig:result}) was the first bright GRB observed by \textit{Fermi}-LAT \citep{080916C}. The afterglow of this GRB was observed in a broad energy band. \cite{Greiner2009} and \cite{Gao2009} performed a multiwavelength analysis and found that the optical to X-ray afterglow emission is within the slow cooling regime, $\nu_{m}$ < $\nu$ < $\nu_{c}$, and the wind environment is preferred over ISM. In our study, we found that the temporal and spectral evolution of high-energy emission is consistent with a closure relation of a cooling regime of $\nu$ > $\nu_{m}$ and $\nu_{c}$ with undetermined environment, and this does not conflict with the multiwavelength result, if $\nu_{c}$ is located between the X-ray and $\gamma$-ray energy bands.

GRB 090510A (orange in Figure~\ref{fig:result}) is the brightest short GRB observed by \textit{Fermi}-LAT in 10 yr, and revealed strong evidence of an additional spectral component for the first time \citep{090510A}. The afterglow of GRB 090510A is modeled by \cite{Pasquale2010} and \cite{,Fraija2016} with a broad energy band from optical to $\gamma$-ray. They found that the simple forward shock model with a constant density medium explains the broad light curve and spectrum successfully. The extrapolation of the X-ray spectrum is consistent with the $\gamma$-ray spectrum, implying that a simple power law explains the two bands. Our result reaches the same conclusion that $\nu_{c}$ should be higher than $\nu_{\rm LAT}$, and the ISM environment is favored. 

GRB 090926A (green in Figure~\ref{fig:result}) is a bright \textit{Fermi}-LAT GRB, showing a spectral break around 1.4 GeV \citep{090926A}. \cite{Cenko2011} and \cite{Gompertz2018} analyzed the afterglow of GRB 090926A with data from optical to $\gamma$-ray, and found that the preferred environment has a wind profile, which disagrees with our result; in our analysis, the best-fit closure relation of GRB 090926A indicates ISM. This conflict may be due to the existence of the spectral break. \cite{Yassine2017} shows that the spectral break of GRB 090926A exists for a long time, and the break energy increases in time. However, in the catalog, the spectrum of the high-energy extended emission is always fitted with a power-law model, which describes the observed data sufficiently well, so that the spectral index from the fit can be an ambiguous value between those of the true spectral indices before and after the break. This may result in the discrepancy between the multiwavelength results and our result.

GRB 110731A (red in Figure~\ref{fig:result}) was simultaneously detected by instruments on board \textit{Fermi} and \textit{Swift} \citep{Ackermann2013}. Also, the rapid follow-up observations by optical telescopes provided multiwavelength data from optical to GeV just after the trigger. \cite{Ackermann2013} concluded that the \textit{Fermi}-LAT extended emission is consistent with emission from a forward shock in a wind medium. In contrast, \cite{Gompertz2018} revisited the multi-wavelength analysis of GRB 110731A and concluded that ISM with the slow cooling condition is favored over wind. Our result is consistent with the result by \cite{Gompertz2018}.

The record-setting GRB 130427A (purple in Figure~\ref{fig:result}) is the brightest GRB with the highest-energy photon (95 GeV) and the longest $\gamma$-ray duration \citep[20 hr;][]{Ackermann2014}. The broadband afterglow of GRB 130427A has been interpreted with the standard forward shock model by many authors \citep[e.g.,][]{Perley2014,Panaitescu2013,Pasquale2016}, and the possible contribution of SSC for explaining the unusual high-energy photon is suggested \citep[e.g.,][]{Fan2013, Liu2013}. \cite{Perley2014} argued that the surrounding environment for GRB 130427A is a wind environment, and \cite{Panaitescu2013} suggested a very tenuous wind. \cite{Pasquale2016} contradicted the argument and explained the evolution of the afterglow of GRB 130427A with a uniform density profile with a 180 day follow-up observation. In these competing models, $\nu_{c}$ is considered to be located below the \textit{Fermi}-LAT energy band, and our classification results in the same conclusion; the best-fit closure relation is $\nu$ $>$ $\nu_{m}$ and $\nu_{c}$ with undetermined environment.
\subsection{Unclassified GRBs} \label{sec:unclassified}
\begin{figure*}[t]
  \centering
  \includegraphics[scale=0.6]{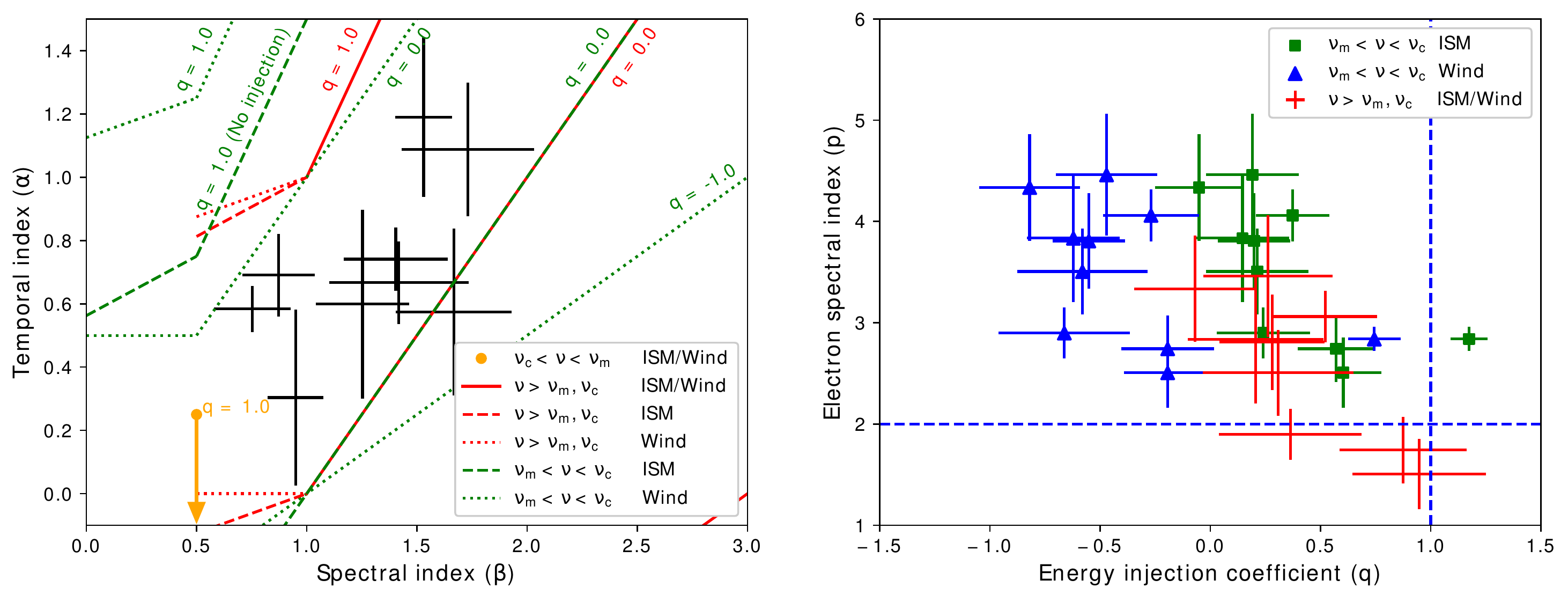}
  \caption{Test of unclassified GRBs with the refreshed shock model, a forward shock with a continuous energy injection. In the left panel, closure relation lines for several energy injection coefficients, \textit{q} = 1, 0 and -1, are plotted with the unclassified GRBs. Note that \textit{q} = 1 is no energy injection. In the right panel, the estimated values of \textit{q} and $\mathit{p}$ for unclassified GRBs are plotted. The red points are when all unclassified GRBs satisfy a closure relation of the fast cooling. The green and blue points are for the slow cooling with ISM and wind, respectively.}
  \label{fig:pq}
\end{figure*}
For unclassified GRBs, we consider modifications to the standard external shock model. First of all, we test an external forward shock model with a continuous energy injection from the central engine during the afterglow phase, the so-called ``refreshed shock model'' \citep{Zhang2001, Zhang2006}. According to this model, the isotropic energy evolves as $E_{iso}$ $\propto$ $t^{1-\textit{q}}$, where \textit{q} is the energy injection coefficient. For values of \textit{q} < 1, there is continuous energy injection from the central engine whereas \textit{q} = 1 is identical to an instantaneous energy injection. The energy injection from a spinning-down newly born magnetar gives \textit{q} = 0 \citep{Dai1998,Zhang2001}. There is no general consensus for the energy injection mechanism and the value of \textit{q}. As the forward external shock is supported by energy injected from the central engine, the temporal index is smaller than in models without energy injections is described as a function of \textit{q} \citep{Zhang2006}, 
\begin{equation}
\begin{aligned}
\alpha (\mathit{p},\mathit{q}) = 
\begin{cases}
\frac{3\mathit{q}-2}{4} &\text{CR1},\\
\frac{(2\mathit{p}-4)+(\mathit{p}+2)\mathit{q}}{4} &\text{CR2},\\
\frac{(2\mathit{p}-6)+(\mathit{p}+3)\mathit{q}}{4} &\text{CR3},\\
\frac{(2\mathit{p}-2)+(\mathit{p}+1)\mathit{q}}{4} &\text{CR4},
\end{cases}
\end{aligned}
\label{eq:qa}
\end{equation}
whereas the spectral index $\beta$ is independent of the energy injection coefficient. As the coefficient \textit{q} decreases (more energy is injected), the temporal index becomes smaller (slower flux decreases). 

Since the true value of \textit{q} for each GRB is unknown, any closure relation (CR2, CR3, and CR4) can be the best-matched closure relation for an unclassified GRB if an appropriate \textit{q} value is chosen. We can calculate a proper value of \textit{q} and corresponding $\mathit{p}$ for each closure relation, assuming that the relation is the best-matched closure relation (equation~\ref{eq:qa}). As shown in Figure~\ref{fig:pq}, the required \textit{q} value varies, and a universal value for \textit{q} is not determined. Rather, we estimate the maximum energy injection coefficient in order to satisfy any one of the closure relations. We find that unclassified GRBs can be explained with values larger than \textit{q} $\sim$ 0.45 on average. The values of $\mathit{p}$ are distributed in the range of 2.5 $\lesssim$ $\mathit{p}$ $\lesssim$ 5, which are slightly higher than those of the classified events. Assuming that all unclassified GRBs meet the same closure relation choosing an appropriate \textit{q} value for each GRB, there may be anticorrelation between \textit{q} and $\mathit{p}$ (see Figure~\ref{fig:pq} right panel), which implies that a larger energy injection results in a softer electron spectrum. We stress that this anticorrelation is only valid under this strong assumption.

In addition, we consider a model describing an external forward shock in the presence of a reverse shock \citep{Meszaros1993, Meszaros1997, Meszaros1999, Sari1999b}. New closure relations can be derived from that model \citep{Kobayashi2000} that have the same form as the closure relations of the refreshed shock model for \textit{q} = 0. In Figure~\ref{fig:pq} (left panel), a few of the unclassified GRBs are consistent with a value of \textit{q} = 0 lines. For those GRBs, this model can explain their evolution of spectrum and light curve.

There are other external shock models for the jet geometry, the structured outflow and the nonuniform surrounding medium \citep[see reviews by ][]{Meszaros2002, Zhang2004, Piran2004, Gao2013}. To test these models for the unclassified GRBs, the detailed characteristics of each GRB should be investigated, which is beyond the scope of this study.

\section{Conclusion} \label{sec:conclusion}
We tested the standard external forward shock model with \textit{Fermi}-LAT GRBs detected from 2008 to 2018. We used a set of closure relations and classified GRBs into the best-matched closure relation. Our main results can be summarized as follows: 
\begin{itemize}
\item First of all, many GRBs satisfy the slow-cooling condition ($\nu_{m}$ < $\nu_{\rm LAT}$ < $\nu_{c}$). We investigated the required physical condition for those GRBs and estimated that the magnetic field energy should occupy a very small fraction of the total energy ($\epsilon_{B}$ $\lesssim$ 10$^{-7}$). About the same number of bursts satisfy a cooling condition of $\nu_{\rm LAT}$ $>$ $\nu_{m}$ and $\nu_{c}$ for which the external density profile cannot be determined.

\item Next, the distribution of the electron spectral index $\mathit{p}$ for the classified GRBs shows median and 1 $\sigma$ of 2.40 and 0.44, respectively. It is interesting that $\mathit{p}$ shows a weak dependence on the cooling regime (KS test = 3.2 $\sigma$). We compared the two cooling regimes: the fast cooling ($\nu$ > $\nu_{m}$ and $\nu_{c}$) and the slow cooling ($\nu_{m}$ < $\nu$ < $\nu_{c}$). The former regime shows a lower median value of the $\mathit{p}$ distribution than that of the later regime, $\mathit{p}_{\rm fast}$ = 2.22 and $\mathit{p}_{\rm slow}$ = 2.61 respectively.

\item Also, we found that a lot of long GRBs meet the closure relation derived in the uniform density profile, which disagrees with the theoretical expectation, but agrees with other observational studies \citep{Yost2003, Schulze2011, Gompertz2018}. 
\end{itemize}
We concluded that the spectrum and temporal evolution of \textit{Fermi}-LAT high-energy (> 100 MeV) extended emission are well explained by the standard external forward shock model, except for a few GRBs. Also, the derived physical conditions such as the electron spectral index, the upper limit of $\epsilon_{B}$, and the surrounding environment do not conflict with those from the X-ray and optical bands.

\acknowledgments
The \textit{Fermi} LAT Collaboration acknowledges generous ongoing support from a number of agencies and institutes that have supported both the development and the operation of the LAT as well as scientific data analysis. These include the National Aeronautics and Space Administration and the Department of Energy in the United States, the Commissariat \`a l'Energie Atomique and the Centre National de la Recherche Scientifique / Institut National de Physique Nucl\'eaire et de Physique des Particules in France, the Agenzia Spaziale Italiana and the Istituto Nazionale di Fisica Nucleare in Italy, the Ministry of Education, Culture, Sports, Science and Technology (MEXT), High Energy Accelerator Research Organization (KEK) and Japan Aerospace Exploration Agency (JAXA) in Japan, and the K.~A.~Wallenberg Foundation, the Swedish Research Council and the Swedish National Space Board in Sweden.
 
Additional support for science analysis during the operations phase is gratefully acknowledged from the Istituto Nazionale di Astrofisica in Italy and the Centre National d'\'Etudes Spatiales in France. This work performed in part under DOE
Contract DE-AC02-76SF00515.\\
\pagebreak


\pagebreak

\bibliographystyle{yahapj}
\bibliography{main.bib}
\appendix
Table~\ref{tab:tot} contains the data and their classification results. For each GRB, the GRB name, the duration category (long or short), and the spectral and temporal indices are listed in 2FLGC. The physical properties corresponding to the best-matched closure relation are listed: the cooling condition, the surrounding environment, and the electron spectral index $\mathit{p}$. The model "UC" indicates an unclassified GRB when none of the eight closure relations can sufficiently account for the observed properties of the GRB (Section~\ref{sec:method}). When the best-matched closure relation does not show a strong preference over some alternative closure relations (Bayes factor < 3), we also include these next-best alternatives in Column ``alt. CR''.
\begin{longtable*}[p]{c c c c c c c c c c c}
\caption{Results of classifying GRBs into the best-matched closure relation}\\
    \hline\hline
& & & & \multicolumn{6}{c}{Best-matched closure relation} \\\cline{5-11.}
GRB Name& Class\footnote{GRBs are classified into long and short duration based upon cut at T\textsubscript{90} = 2.} & $\beta$ & $\alpha$ & Model & Cooling Regime & Environment & $\mathit{p}$\textsubscript{$\beta$} &  $\mathit{p}$\textsubscript{$\alpha$} & p\textsubscript{ave} & Alt. CR\footnote{The Bayes factor of the best-matched closure relation over these alternatives is less than 3}\\\hline
080916C & L & 1.16$\pm$0.11 & 1.13$\pm$0.12 & CR2 &$\nu$ > $\nu_{m}$, $\nu_{c}$&ISM/Wind& 2.3$\pm$0.2 & 2.2$\pm$0.2 & 2.2$\pm$0.1\\
081009 & L & 0.79$\pm$0.23 & 0.88$\pm$0.24 & CR3 &$\nu_{m}$ < $\nu$ < $\nu_{c}$&ISM& 2.6$\pm$0.5 & 2.2$\pm$0.3 & 2.3$\pm$0.3 & 5, 7\\
090323 & L & 1.15$\pm$0.16 & 1.09$\pm$0.13 & CR2 &$\nu$ > $\nu_{m}$, $\nu_{c}$&ISM/Wind& 2.3$\pm$0.3 & 2.1$\pm$0.2 & 2.2$\pm$0.2\\
090328A & L & 1.18$\pm$0.14 & 0.99$\pm$0.09 & CR2 &$\nu$ > $\nu_{m}$, $\nu_{c}$&ISM/Wind& 2.4$\pm$0.3 & 2.0$\pm$0.1 & 2.1$\pm$0.1\\
090510 & S & 1.15$\pm$0.08 & 1.81$\pm$0.08 & CR3 &$\nu_{m}$ < $\nu$ < $\nu_{c}$&ISM& 3.3$\pm$0.2 & 3.4$\pm$0.1 & 3.4$\pm$0.1\\
090626 & L & 1.00$\pm$0.21 & 0.94$\pm$0.22 & CR2 &$\nu$ > $\nu_{m}$, $\nu_{c}$&ISM/Wind& 2.0$\pm$0.4 & 1.9$\pm$0.3 & 2.0$\pm$0.2 & 3, 5, 7\\
090902B & L & 0.92$\pm$0.06 & 1.63$\pm$0.08 & UC & & & \\
090926A & L & 0.86$\pm$0.07 & 1.39$\pm$0.08 & CR3 &$\nu_{m}$ < $\nu$ < $\nu_{c}$&ISM& 2.7$\pm$0.1 & 2.9$\pm$0.1 & 2.8$\pm$0.1\\
091003 & L & 0.84$\pm$0.18 & 0.91$\pm$0.22 & CR3 &$\nu_{m}$ < $\nu$ < $\nu_{c}$&ISM& 2.7$\pm$0.4 & 2.2$\pm$0.3 & 2.4$\pm$0.2 & 2, 5, 7\\
091031 & L & 0.81$\pm$0.23 & 1.26$\pm$0.21 & CR3 &$\nu_{m}$ < $\nu$ < $\nu_{c}$&ISM& 2.6$\pm$0.5 & 2.7$\pm$0.3 & 2.7$\pm$0.2 & 4\\
100116A & L & 0.59$\pm$0.18 & 2.70$\pm$0.19 & UC & & & \\
100414A & L & 0.79$\pm$0.12 & 1.27$\pm$0.13 & CR3 &$\nu_{m}$ < $\nu$ < $\nu_{c}$&ISM& 2.6$\pm$0.2 & 2.7$\pm$0.2 & 2.7$\pm$0.1\\
100511A & L & 0.75$\pm$0.17 & 0.58$\pm$0.07 & UC & & & \\
110428A & L & 0.95$\pm$0.25 & 0.98$\pm$0.11 & CR2 &$\nu$ > $\nu_{m}$, $\nu_{c}$&ISM/Wind& 1.9$\pm$0.5 & 2.0$\pm$0.2 & 2.0$\pm$0.1 & 3, 5, 7\\
110625A & L & 1.67$\pm$0.26 & 0.57$\pm$0.26 & UC & & & \\
110731A & L & 1.00$\pm$0.20 & 1.53$\pm$0.12 & CR3 &$\nu_{m}$ < $\nu$ < $\nu_{c}$&ISM& 3.0$\pm$0.4 & 3.0$\pm$0.2 & 3.0$\pm$0.1 & 4\\
120526A & L & 0.87$\pm$0.16 & 0.69$\pm$0.13 & UC & & & \\
120624B & L & 1.53$\pm$0.13 & 1.19$\pm$0.25 & UC & & & \\
120709A & L & 1.42$\pm$0.32 & 0.67$\pm$0.13 & UC & & & \\
120711A & L & 1.08$\pm$0.17 & 1.63$\pm$0.24 & CR3 &$\nu_{m}$ < $\nu$ < $\nu_{c}$&ISM& 3.2$\pm$0.3 & 3.2$\pm$0.3 & 3.2$\pm$0.2 & 2, 4\\
120911B & L & 1.33$\pm$0.30 & 1.31$\pm$0.20 & CR2 &$\nu$ > $\nu_{m}$, $\nu_{c}$&ISM/Wind& 2.7$\pm$0.6 & 2.4$\pm$0.3 & 2.5$\pm$0.2 & 3\\
130325A & L & 0.69$\pm$0.28 & 0.13$\pm$0.32 & CR1 &$\nu_{c}$ < $\nu$ < $\nu_{m}$&ISM/Wind& & & \\
130327B & L & 0.76$\pm$0.13 & 1.56$\pm$0.16 & CR4 &$\nu_{m}$ < $\nu$ < $\nu_{c}$&Wind& 2.5$\pm$0.3 & 2.4$\pm$0.2 & 2.5$\pm$0.2\\
130427A & L & 1.12$\pm$0.06 & 1.24$\pm$0.06 & CR2 &$\nu$ > $\nu_{m}$, $\nu_{c}$&ISM/Wind& 2.2$\pm$0.1 & 2.3$\pm$0.1 & 2.3$\pm$0.1\\
130502B & L & 0.99$\pm$0.14 & 1.44$\pm$0.06 & CR3 &$\nu_{m}$ < $\nu$ < $\nu_{c}$&ISM& 3.0$\pm$0.3 & 2.9$\pm$0.1 & 2.9$\pm$0.1\\
130504C & L & 0.86$\pm$0.21 & 0.77$\pm$0.06 & CR3 &$\nu_{m}$ < $\nu$ < $\nu_{c}$&ISM& 2.7$\pm$0.4 & 2.0$\pm$0.1 & 2.0$\pm$0.1\\
130518A & L & 1.73$\pm$0.30 & 1.09$\pm$0.21 & UC & & & \\
130606B & L & 0.74$\pm$0.20 & 0.67$\pm$0.20 & CR3 &$\nu_{m}$ < $\nu$ < $\nu_{c}$&ISM& 2.5$\pm$0.4 & 1.9$\pm$0.3 & 2.1$\pm$0.2 & \\\hline
130821A & L & 1.35$\pm$0.25 & 0.99$\pm$0.14 & CR2 &$\nu$ > $\nu_{m}$, $\nu_{c}$&ISM/Wind& 2.7$\pm$0.5 & 2.0$\pm$0.2 & 2.1$\pm$0.2\\
131014A & L & 0.97$\pm$0.25 & 0.82$\pm$0.17 & CR5 &$\nu$ > $\nu_{m}$, $\nu_{c}$&ISM/Wind& 1.9$\pm$0.5 & 1.0$\pm$0.9 & 1.7$\pm$0.4 & 2, 3, 7\\
131029A & L & 1.27$\pm$0.33 & 1.11$\pm$0.17 & CR2 &$\nu$ > $\nu_{m}$, $\nu_{c}$&ISM/Wind& 2.5$\pm$0.7 & 2.1$\pm$0.2 & 2.2$\pm$0.2 & 3\\
131108A & L & 1.79$\pm$0.27 & 1.51$\pm$0.15 & CR2 &$\nu$ > $\nu_{m}$, $\nu_{c}$&ISM/Wind& 3.6$\pm$0.5 & 2.7$\pm$0.2 & 2.8$\pm$0.2\\
131231A & L & 0.63$\pm$0.12 & 1.03$\pm$0.21 & CR3 &$\nu_{m}$ < $\nu$ < $\nu_{c}$&ISM& 2.3$\pm$0.2 & 2.4$\pm$0.3 & 2.3$\pm$0.2\\
140206B & L & 0.95$\pm$0.13 & 0.30$\pm$0.28 & UC & & & \\
140523A & L & 1.03$\pm$0.19 & 0.96$\pm$0.14 & CR2 &$\nu$ > $\nu_{m}$, $\nu_{c}$&ISM/Wind& 2.1$\pm$0.4 & 1.9$\pm$0.2 & 2.0$\pm$0.2 & 5, 7\\
140810A & L & 0.55$\pm$0.21 & 0.82$\pm$0.19 & CR3 &$\nu_{m}$ < $\nu$ < $\nu_{c}$&ISM& 2.1$\pm$0.4 & 2.1$\pm$0.3 & 2.1$\pm$0.2 & 5\\
141028A & L & 1.01$\pm$0.25 & 0.97$\pm$0.03 & CR5 &$\nu$ > $\nu_{m}$, $\nu_{c}$&ISM/Wind& 2.0$\pm$0.5 & 1.8$\pm$0.2 & 1.9$\pm$0.2 & 2, 3, 7\\
141207A & L & 0.80$\pm$0.30 & 1.88$\pm$0.03 & CR4 &$\nu_{m}$ < $\nu$ < $\nu_{c}$&Wind& 2.6$\pm$0.6 & 2.8$\pm$0.0 & 2.8$\pm$0.0 & 3\\
141222A & L & 1.10$\pm$0.33 & 1.33$\pm$0.40 & CR2 &$\nu$ > $\nu_{m}$, $\nu_{c}$&ISM/Wind& 2.2$\pm$0.7 & 2.4$\pm$0.5 & 2.3$\pm$0.4 & 3, 4\\
150523A & L & 0.78$\pm$0.16 & 1.03$\pm$0.25 & CR3 &$\nu_{m}$ < $\nu$ < $\nu_{c}$&ISM& 2.6$\pm$0.3 & 2.4$\pm$0.3 & 2.5$\pm$0.2\\
150627A & L & 0.68$\pm$0.12 & 0.93$\pm$0.20 & CR3 &$\nu_{m}$ < $\nu$ < $\nu_{c}$&ISM& 2.4$\pm$0.2 & 2.2$\pm$0.3 & 2.3$\pm$0.2\\
150902A & L & 1.06$\pm$0.20 & 1.05$\pm$0.18 & CR2 &$\nu$ > $\nu_{m}$, $\nu_{c}$&ISM/Wind& 2.1$\pm$0.4 & 2.1$\pm$0.2 & 2.1$\pm$0.2\\
160325A & L & 1.40$\pm$0.24 & 0.74$\pm$0.10 & UC & & & \\
160509A & L & 1.10$\pm$0.30 & 1.13$\pm$0.11 & CR2 &$\nu$ > $\nu_{m}$, $\nu_{c}$&ISM/Wind& 2.2$\pm$0.6 & 2.2$\pm$0.2 & 2.2$\pm$0.1 & 3\\
160521B & L & 0.41$\pm$0.26 & 1.35$\pm$0.20 & CR4 &$\nu_{m}$ < $\nu$ < $\nu_{c}$&Wind& 1.8$\pm$0.5 & 2.1$\pm$0.3 & 2.1$\pm$0.2\\
160623A & L & 0.98$\pm$0.11 & 1.25$\pm$0.09 & CR3 &$\nu_{m}$ < $\nu$ < $\nu_{c}$&ISM& 3.0$\pm$0.2 & 2.7$\pm$0.1 & 2.7$\pm$0.1 & 2\\
160625B & L & 0.75$\pm$0.28 & 2.24$\pm$0.28 & CR4 &$\nu_{m}$ < $\nu$ < $\nu_{c}$&Wind& 2.5$\pm$0.6 & 3.3$\pm$0.4 & 3.1$\pm$0.3\\
160816A & L & 1.16$\pm$0.18 & 1.25$\pm$0.13 & CR2 &$\nu$ > $\nu_{m}$, $\nu_{c}$&ISM/Wind& 2.3$\pm$0.4 & 2.3$\pm$0.2 & 2.3$\pm$0.2\\
160821A & L & 0.75$\pm$0.17 & 1.15$\pm$0.10 & CR3 &$\nu_{m}$ < $\nu$ < $\nu_{c}$&ISM& 2.5$\pm$0.3 & 2.5$\pm$0.1 & 2.5$\pm$0.1\\
160905A & L & 0.84$\pm$0.19 & 1.15$\pm$0.28 & CR3 &$\nu_{m}$ < $\nu$ < $\nu_{c}$&ISM& 2.7$\pm$0.4 & 2.5$\pm$0.4 & 2.6$\pm$0.3\\
170214A & L & 1.33$\pm$0.11 & 1.73$\pm$0.26 & CR2 &$\nu$ > $\nu_{m}$, $\nu_{c}$&ISM/Wind& 2.7$\pm$0.2 & 3.0$\pm$0.3 & 2.7$\pm$0.2 & 3\\
170405A & L & 1.51$\pm$0.33 & 1.27$\pm$0.01 & CR2 &$\nu$ > $\nu_{m}$, $\nu_{c}$&ISM/Wind& 3.0$\pm$0.7 & 2.4$\pm$0.0 & 2.4$\pm$0.0\\
170808B & L & 1.15$\pm$0.26 & 1.01$\pm$0.22 & CR2 &$\nu$ > $\nu_{m}$, $\nu_{c}$&ISM/Wind& 2.3$\pm$0.5 & 2.0$\pm$0.3 & 2.1$\pm$0.3\\
170906A & L & 1.06$\pm$0.14 & 0.83$\pm$0.12 & CR5 &$\nu$ > $\nu_{m}$, $\nu_{c}$&ISM/Wind& 2.1$\pm$0.3 & 1.1$\pm$0.7 & 2.0$\pm$0.3 & 2, 7\\
171010A & L & 1.05$\pm$0.13 & 1.32$\pm$0.16 & CR2 &$\nu$ > $\nu_{m}$, $\nu_{c}$&ISM/Wind& 2.1$\pm$0.3 & 2.4$\pm$0.2 & 2.3$\pm$0.2 & 3\\
171120A & L & 1.25$\pm$0.21 & 0.60$\pm$0.30 & UC & & & \\
180210A & L & 0.74$\pm$0.14 & 1.03$\pm$0.19 & CR3 &$\nu_{m}$ < $\nu$ < $\nu_{c}$&ISM& 2.5$\pm$0.3 & 2.4$\pm$0.2 & 2.4$\pm$0.2\\
180703A & L & 1.38$\pm$0.33 & 0.85$\pm$0.15 & CR2 &$\nu$ > $\nu_{m}$, $\nu_{c}$&ISM/Wind& 2.8$\pm$0.7 & 1.8$\pm$0.2 & 1.9$\pm$0.2 & 3, 5\\
180720B & L & 1.15$\pm$0.10 & 1.88$\pm$0.15 & CR3 &$\nu_{m}$ < $\nu$ < $\nu_{c}$&ISM& 3.3$\pm$0.2 & 3.5$\pm$0.2 & 3.4$\pm$0.1 & 4\\\hline\hline

\label{tab:tot} 
\end{longtable*}
\end{document}